\documentclass[amsmath,amssymb,aps,prl,showpacs,lengthcheck]{revtex4}

\usepackage{graphicx}

\begin{document}

\title{Constraints on the symmetry energy and on neutron skins 
from the pygmy resonances in $^{68}$Ni and $^{132}$Sn}

\author{Andrea Carbone$^1$}
\author{Gianluca Col\`o$^{1,2}$}
\author{Angela Bracco$^{1,2}$}
\author{Li-Gang Cao$^{1,2,3,4}$}
\author{Pier Francesco Bortignon$^{1,2}$}
\author{Franco Camera$^{1,2}$}
\author{Oliver Wieland$^2$}

\affiliation{$^1$ Dipartimento di Fisica, Universit\`a degli 
Studi di Milano, via Celoria 16, 20133 Milano, Italy}
\affiliation{$^2$ INFN, Sezione di Milano, via Celoria 16, 
20133 Milano, Italy}
\affiliation{$^3$ Institute of Modern Physics, Chinese Academy of
Science, Lanzhou 730000, P.R. China}
\affiliation{$^4$ Center of Theoretical Nuclear Physics,
National Laboratory of Heavy Ion Accelerator of Lanzhou, Lanzhou
730000, P.R. China}

\date{\today}

\begin{abstract}
Correlations between the behavior of the nuclear symmetry
energy, the neutron skins, and the percentage of energy-weighted sum rule
(EWSR) exhausted by the Pygmy Dipole Resonance (PDR) in $^{68}$Ni 
and $^{132}$Sn have been investigated by using different 
Random Phase Approximation (RPA) models for the dipole response, based on  
a representative set of Skyrme effective forces plus meson-exchange
effective Lagrangians. A comparison 
with the experimental data has allowed us to constrain the value of 
the derivative of the symmetry energy at saturation. 
The neutron skin radius is deduced under this constraint.
\end{abstract}


\maketitle

One of the interesting problems presently receiving particular
attention, is that of the size of the neutron root-mean-square (r.m.s.) 
radius in neutron-rich nuclei. In fact, this quantity is
related to the isospin-dependent part of the nuclear equation 
of state (EOS) which, in turn, has relevant implications 
for the description of neutron stars. At present, there is
an enormous effort aimed at determining the parameters that
govern the asymmetric matter EOS, using both experimental
and theoretical tools. Review papers have been devoted to
this topic \cite{Steiner,Li}. 

The energy per particle in a nuclear system characterized by a total density
$\rho$ (sum of the neutron and proton densities $\rho_n$ and $\rho_p$), 
and by a local 
asymmetry $\delta \equiv \left( \rho_n-\rho_p
\right)/\rho$, is usually written as
\begin{equation}
\frac{E}{A}\left( \rho,\delta \right) = \frac{E}{A}\left( \rho, 
\delta=0 \right) + S(\rho)\delta^2.
\end{equation}
Odd powers of $\delta$ are forbidden by the isospin symmetry and
the term proportional to $\delta^4$ is found to be negligible. The above equation
defines the so-called symmetry energy $S(\rho)$. Determining values of the
symmetry energy at various densities of interest for nuclear
structure, nuclear reactions, and astrophysics, is one of the
great challenges for the physics community. 

Information on the symmetry energy can be obtained from various sources,
none of them being so far conclusive by itself. A direct 
correlation between the neutron skin thickness $\Delta R$ 
and the derivative of the symmetry energy at saturation, has been found 
in Refs. \cite{Brown,Furnstahl}. The derivative of the 
symmetry energy at saturation is related to the 
widely used ``slope'' parameter $L$ by 
\begin{equation}
S^\prime(\rho)\vert_{\rho=\rho_0} = \frac{L}{3\rho_0}.
\label{L}
\end{equation}
The symmetry energy at saturation, $S(\rho_0)$, 
is denoted by $a_4$ or $J$: we shall use the symbol $J$ in what
follows. No measurement of the 
neutron skin is available which is accurate enough to 
constrain the slope parameter $L$. 
The properties of the isovector Giant Dipole Resonance (IVGDR) 
\cite{Trippa}, of the low-lying electric dipole excitation 
(the so-called Pygmy Dipole Resonance, PDR) \cite{Klimkiewicz}, 
and of the charge-exchange spin-dipole strength \cite{Yako} have 
been suggested as constraints. In addition, by means of heavy-ion 
collisions the symmetry energy has also been probed at 
subsaturation densities (0.4 $\leq$ $\rho/\rho_0$$\leq$ 1.2).
In Ref. \cite{Chen}, isospin diffusion data from the collision 
between $^{112}$Sn and $^{124}$Sn have been analyzed
using a transport model in which the momentum-dependent symmetry
potential enters as one of the main ingredients. 
The same data, together with the double ratios of neutron and proton 
energy spectra, have been analyzed within a different kind of
transport model in Ref. \cite{Tsang}. We 
should also mention that another analysis of
isoscaling data has been reported in  Ref. \cite{Shetty}.
Finally, in the work reported in Ref. \cite{Warda} a range of values
for $L$ is inferred from the analysis of data of radii
from antiprotonic atoms. 

One of the motivations of the present work lies in the
consideration that the values of $L$ extracted from the PDR
in $^{132}$Sn (between $\approx$ 30 and 60 MeV) \cite{Klimkiewicz} 
and from the GDR in $^{208}$Pb \cite{footnote1}, 
are smaller than those deduced 
from the analysis of the heavy-ion collisions. More precisely, 
the values of $L$ deduced with the two different approaches 
overlap only in a small interval (cf., e.g., Fig. 3 
of \cite{Tsang}). We would like to pursue in this paper 
an analysis which is more general than the one performed 
in Ref. \cite{Klimkiewicz}, by considering PDRs in two different mass
regions and a variety of theoretical models, both nonrelativistic
and relativistic. Both classes of mean-field models are
successful in describing the nuclear ground states and of many of the excited
states (for a review, see Ref. \cite{Bender}). 
Our goal is to see whether consistency among 
different ways of extracting the slope parameter $L$ can be 
achieved: as a result, one should also expect to be able to 
better pin down the values of the neutron skin radii. 

Progress in the study of the density 
dependence of the symmetry energy and the neutron radii 
through the PDR requires more work in two directions.
The first is to have more data, particularly for unstable neutron
rich-nuclei characterized by a sizeable dipole strength in the low-energy 
region. The second 
is a more comprehensive theoretical analysis 
of the PDRs based on different calculations both of nonrelativistic 
and relativistic RPA types. In this paper this is realized 
by taking advantage of a recent experimental datum on the PDR
in the neutron-rich $^{68}$Ni nucleus \cite{Wieland}. In particular, the
present analysis has aimed (i) at finding evidence of a correlation
between the values of $L$ and the EWSR exhausted by the PDRs when they 
are calculated using many Skyrme parameter sets and covariant
effective Lagrangians, (ii) at inferring values for the neutron skin radii
of $^{68}$Ni, $^{132}$Sn and $^{208}$Pb, and finally (iii) at 
comparing our deduced value of $L$ with the other ones existing in
the recent literature. 

The first step consisted in finding correlations between 
the properties of the PDR and the symmetry energy. 
Random Phase Approximation (RPA) 
calculations of the dipole strength have been carried out. Our 
implementation based on nonrelativistic Skyrme forces is fully 
self-consistent and discussed, e.g., in Ref. \cite{comex3}. The 
Hartree-Fock (HF) equations are solved in a radial mesh extending up to 
$\approx$4 times the nuclear radius. The continuum is discretized by 
using box boundary conditions. The model space is large enough so that 
the well-known double commutator EWSR is exhausted by at least 96\%. 
We have employed 19 different Skyrme sets which can be said to 
constitute a quite representative ensemble. All of them have
an associated value of the nuclear incompressibility $K_\infty$
lying in the interval 210-270 MeV \cite{K}. 
We do not provide here the original
references in which the parameter sets have been introduced:
they can be found in \cite{Trippa,Stone}. We have 
checked that our results do not change appreciably if we 
take out, or add, a few Skyrme parameter sets to our 
ensemble. The relativistic calculations are based on the well-known 
relativistic mean field (RMF) theory 
plus the self-consistent relativistic RPA (RRPA) as described in 
Refs. \cite{Ma02,Ring01}. We have employed 7 different parametrizations
for the non-linear, meson-exchange effective Lagrangian. In 
the calculations, box boundary conditions are used: the box radius 
is set at 30 fm and the radial mesh is 0.1 fm. The model 
spaces for particle-hole and antiparticle-hole are cut at 
energy $E_{\rm cutoff}$ = 1039 MeV and -939 MeV, 
respectively. The references for the seven 
parameter sets can be found in \cite{Cao,Sulaksono}. 

\begin{figure}[!htbp]
\begin{center}
\includegraphics[width=0.4\textwidth]{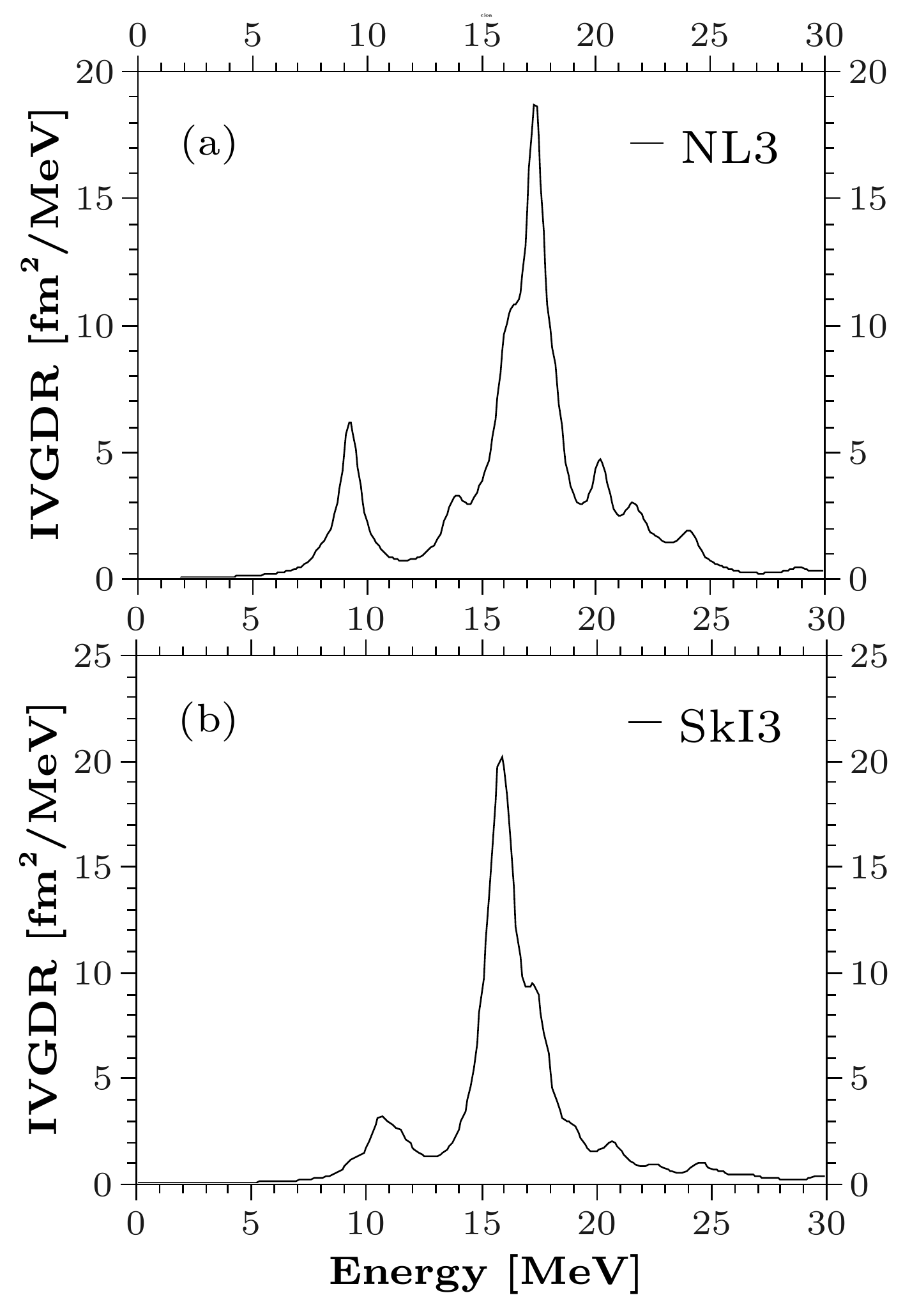}
\end{center}
\caption{Two typical dipole strength functions calculated in the 
nucleus $^{68}$Ni. A nonrelativistic and a relativistic example are shown 
in panels (a) and (b) in which, respectively, the Skyrme
force SkI3 and the NL3 parametrization of the effective RMF Lagrangian have
been used. The
sharp RPA peaks are averaged by using Lorentzian functions having 1 MeV width.}
\label{fig1}
\end{figure}

We have found a rather good correlation between the parameter
$L$ and the percentage of EWSR associated with the PDR. In the
theoretical calculations, we consider the whole part of the low-energy
region where the strength is not negligible. In the nucleus 
$^{68}$Ni the PDR is associated, 
as a rule, with a well-defined peak between 9 and 11 MeV, to be
compared with the experimental finding of \cite{Wieland}, that is, 
11 MeV. In few cases the strength is more fragmented and/or at lower energy. 
We display two typical dipole strength distributions in 
Fig. \ref{fig1}: the separation between PDR and GDR regions
looks quite clear. 
In the nucleus $^{132}$Sn, the peak of the PDR 
is between 7.5 and 9.5 MeV. The experimental peak energy
is 9.8 MeV \cite{Klimkiewicz}. The percentages of EWSR are 
defined in this work with respect to the classical Thomas-Reihe-Kuhn 
(TRK) value, and vary between 1\% and 10\%. In general, the 
relativistic Lagrangians provide larger values for this 
latter quantity. The PDR energies they provide in 
the case of $^{132}$Sn are also about 1 MeV lower than the
experimental value: therefore, trying to constrain the symmetry
energy by using the correlation between the PDR energy and
the value of $S$ at $\rho$=0.1 fm$^{-3}$ (plus the experimental
datum) was attempted in \cite{Cao}, but it was only possible by
means of extrapolation.

\begin{figure}[!htbp]
\begin{center}
\includegraphics[width=1.0\textwidth]{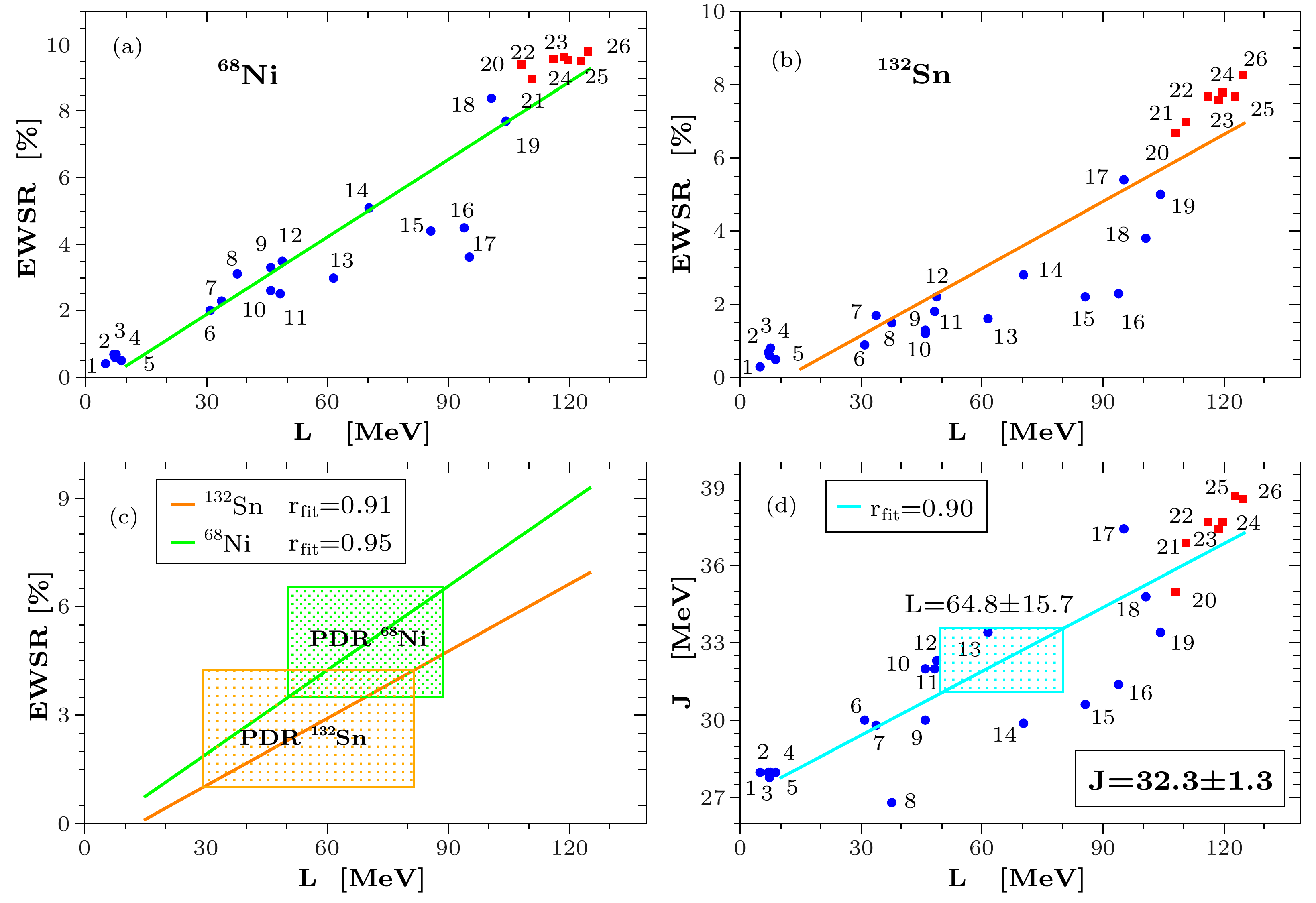}
\end{center}
\caption{(Color online) In panels (a) and (b), the correlation 
between $L$ and the percentage of TRK sum
rule exhausted by the PDR, respectively in $^{68}$Ni and $^{132}$Sn, is
displayed. The computed data points are 
labelled, here and in what follows, by numbers. The correspondence with
the parameter sets used is: 1=v090, 2=MSk3, 3=BSk1, 4=v110, 5=v100, 
6=SkT6, 7=SkT9, 8=SGII, 9=SkM*, 10=SLy4, 11=SLy5, 12=SLy230a, 13=LNS, 
14=SkMP, 15=SkRs, 16=SkGs, 17=SK255, 18=SkI3, 19=SkI2, 20=NLC, 21=TM1, 
22=PK1, 23=NL3, 24=NLBA, 25=NL3+, 26=NLE. The straight lines correspond 
to the results of the fits. In panel (c) we show the 
same straight lines displayed in (a) and (b), together with the correlation
coefficient $r$ and the constraints from experiments \cite{Klimkiewicz,Wieland}. 
In panel (d) the correlation between $L$ and $J$ is shown. The box 
corresponds to the value of $L$ deduced from the weighted average of 
the two values extracted from $^{68}$Ni and $^{132}$Sn.}
\label{fig2}
\end{figure}

In the upper part of Fig. \ref{fig2} the correlation between the 
percentage of the EWSR and $L$ is shown for both nuclei
$^{68}$Ni and $^{132}$Sn. The straight lines correspond to 
linear fits. We have considered the measured values of 
the EWSR percentage, and deduced a range of acceptable values for $L$, 
by taking care both of the experimental error and of the
error associated with the fit (the latter being almost negligible with
respect to the former). Our results are more general than those presented in 
Ref. \cite{Klimkiewicz} since we consider two different
nuclei and many different mean field models. Although 
we do not include in our analysis all classes of mean field models, 
we try nonetheless to avoid, as much as we can, possible 
sources of bias since we avoid restricting to Skyrme sets 
fitted by the same group with the same protocol. In fact, 
our sets span a broad range of possible values associated with 
nuclear matter quantities.

In the case of $^{68}$Ni the measured value of the EWSR percentage is 
5\%$\pm$1.5\%. The error includes the uncertainty related to the 
quantities used to deduce the number from the measurement. 
It should be noted that the dominating uncertainty (still within 
30\% of the average value) is that related to the choice of the 
level density value entering the evaluation of the branching for 
gamma emission. We have used different level 
densities obtained by means of either
a Shell Model Monte Carlo (SMMC) calculation for this 
nucleus \cite{Alhassid}, or global Hartree-Fock-Bogoliubov 
(HFB) calculations \cite{BSk14,MSk7}: the largest span goes 
from 3.5\% obtained using \cite{BSk14} to 6.5\% using \cite{MSk7}.

Our result is that the slope parameter $L$ is constrained to
be in the interval 50.3-89.4 MeV or 29.0-82.0 MeV, if we use 
either the $^{68}$Ni results, or the $^{132}$Sn results (cf. 
the lower left panel of Fig. \ref{fig2}). The 
weighted average, $L$ = 64.8$\pm$15.7 MeV, is displayed in 
the lower right panel of Fig. \ref{fig2} (it corresponds to
the shaded box). In this panel, the correlation of $J$ and $L$ 
is provided, so that we can deduce our best value of $J$ 
which is 32.3$\pm$1.3 MeV. This value is in very good agreement 
with the value 32.0$\pm$1.8 MeV which 
is reported in Ref. \cite{Klimkiewicz}. The parametrizations
of $S(\rho)$ found in Refs. \cite{Chen,Shetty} lead 
to $J$ = 31.6 MeV. Moreover, our result for $J$ overlaps 
well with the ranges obtained in Refs. \cite{Tsang} (30.2-33.8 MeV) 
and \cite{Danielewicz2} (31.5-33.5 MeV) (cf. 
also \cite{Danielewicz1}). From the theoretical point of view, we
can consider very satisfactory that our result for $L$ coincides
almost exactly with the value of 66.5 MeV extracted from Bruckner-Hartree-Fock 
(BHF) calculations in uniform matter that employ realistic two-body and three-body
forces \cite{Vidana}. 

\begin{figure}[!htbp]
\begin{center}
\includegraphics[width=1.0\textwidth]{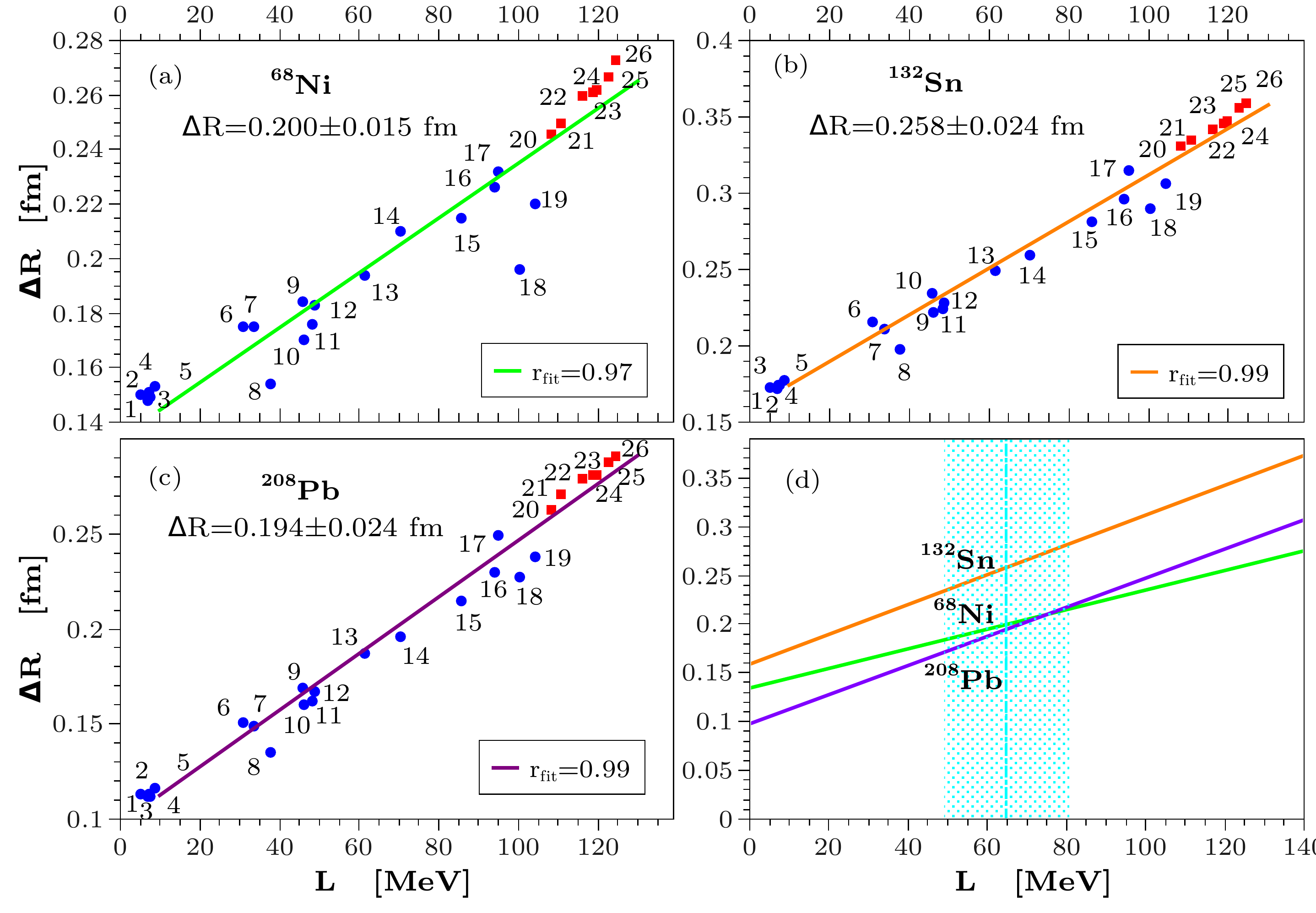}
\end{center}
\caption{(Color online) Panels (a), (b) and (c) display 
the correlations between 
the neutron skin thickness $\Delta R$ and the slope parameter $L$, 
in the case of the three nuclei analyzed in this work. The convention 
is the same as in the previous figure. Under the constraint for $L$ 
emerging from our analysis [shaded area in panel (d)],  
the values displayed for the neutron skin thickness in the 
three nuclei are obtained.}
\label{fig3}
\end{figure}

The next step is to use the $L$ value obtained from the PDR
computed data points in $^{68}$Ni and $^{132}$Sn in order to deduce the
neutron skin thickness $\Delta R$. First, one can note that
the correlation between $L$ and $\Delta R$, when the two
quantities are calculated using the models already described, 
is quite good (cf. Fig. \ref{fig3}). If one imposes the value 
of $L$ to be in the interval 64.8$\pm$15.7 MeV, 
one obtains for the skin thickness $\Delta R$=0.200$\pm$0.015 fm for  
$^{68}$Ni, $\Delta R$=0.258$\pm$0.024 fm for $^{132}$Sn, and 
$\Delta R$=0.194$\pm$0.024 fm for $^{208}$Pb. These numbers are 
stable if one tries to constrain them
by using the $L$ value from $^{68}$Ni only, or $^{132}$Sn only, 
instead of the weighted average. It should also be noted that 
the values associated with $\Delta R$, both for $^{132}$Sn and $^{208}$Pb,  
are in good agreement with the results reported in Ref. \cite{Klimkiewicz}. 
This gives us further confidence on the value of the neutron skin 
of $^{68}$Ni which is determined for the first time 
through the present analysis. We should recall that the
possibility to extract $\Delta R$ directly from 
measurements of the spin-dipole strength has been 
discussed \cite{Krasznahorkay}. For a thorough discussion of the
exensive literature appeared in previous decades on this subject we
refer the reader to Ref. \cite{Harakeh}.  

\begin{figure}[!htbp]
\begin{center}
\includegraphics[width=0.4\textwidth]{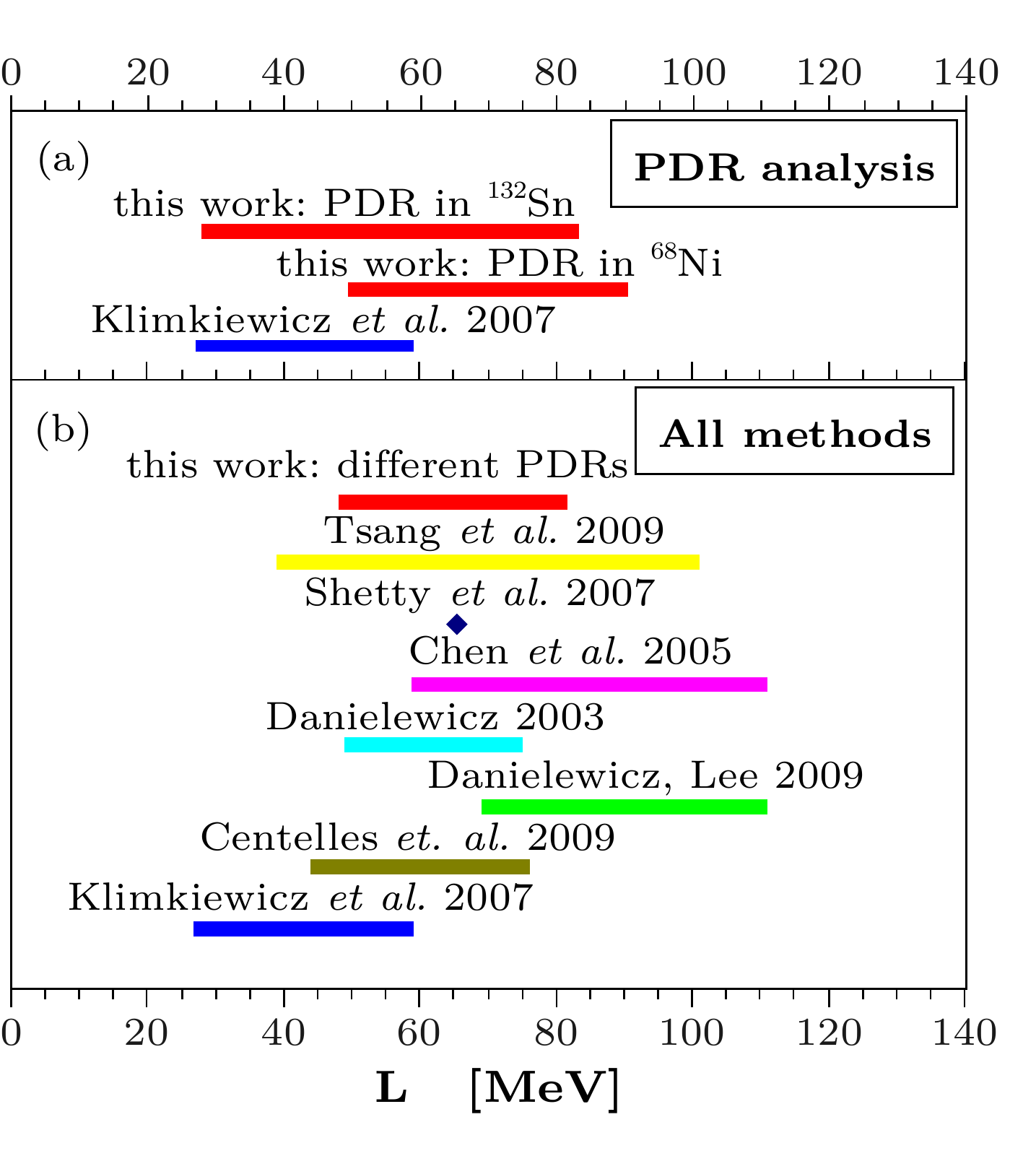}
\end{center}
\caption{(Color online) Comparison between the values of $L$ extracted 
in the present work, and those from the existing literature. In panel 
(a), we compare our separate results from the PDRs of $^{68}$Ni and 
$^{132}$Sn with the result of Klimkiewicz {\it et al.} \cite{Klimkiewicz}. 
In panel (b), we compare with values extracted from completely different 
kind of analysis: Tsang {\it et al.} \cite{Tsang}, Shetty {\it et al.} 
\cite{Shetty}, Chen {\it et al.} \cite{Chen}, Danielewicz \cite{Danielewicz1}, 
Danielewicz and Lee \cite{Danielewicz2}, Centelles {\it et al.} \cite{Warda}, 
and Klimkiewicz {\it et al.} \cite{Klimkiewicz}.}
\label{fig4}
\end{figure}

In Fig. \ref{fig4} we show the comparison of the values of $L$ 
found in our analysis with those found with other analysis and/or
other methods. The main point is that the result for $L$ extracted
from the PDR in $^{132}$Sn is compatible with
the one from Ref. \cite{Klimkiewicz}; however, combining in our
analysis the two PDRs of both $^{132}$Sn and $^{68}$Ni, we are
able to shift the range of $L$ to larger values and to reduce 
the uncertainty. This solves, to a good extent, the problem
that the result from Ref. \cite{Klimkiewicz} was not overlapping
significantly with the results obtained by the different 
analysis of heavy-ion collisions. In the lower panel of 
Fig. \ref{fig4}, one can see that our present finding has
a remarkable overlap with the results of most of the other
proposed methods to extract $L$, that involve not only 
different methodologies but also very different observables. 
We can conclude that our more general analysis of the extraction
of the slope parameter $L$ from the PDR is able to provide a firm
result. Another important side result of our work is that we are
able for the first time to propose a value for the neutron
skin thickness of the neutron-rich $^{68}$Ni isotope. More PDR data 
in other mass regions and/or in long isotopic chains are desirable to increase the
predictive power of our procedure. This could lead to determine
quite accurately quantities such as the neutron radii, and the 
parameters governing the density dependence of the symmetry energy, 
that are fundamental for nuclear physics and for their 
implications in the study of neutron stars.

C.L. acknowledges the support of the UniAMO grant 
provided by Fondazione Cariplo and
Universit\`a degli Studi which has allowed his stay in
Milano, and the support of the National Science Foundation
of China under Grant Nos. 10875150.

\newpage

\begin{thebibliography}{00}

\bibitem{Steiner} A.~W.~Steiner {\em et al.}, Phys. Rep. 411, 
325 (2005).
\bibitem{Li} B.~A.~Li {\em et al.}, Phys. Rep. 464, 113 (2008).
\bibitem{Brown} B.~A.~Brown, Phys. Rev. Lett. 85, 5296 (2000); 
S.~Typel and B.~A.~Brown, Phys. Rev. C64, 027302(R) (2001). 
\bibitem{Furnstahl} R.~J.~Furnstahl, Nucl. Phys. A706, 85 (2002). 
\bibitem{Trippa} L.~Trippa {\em et al.}, Phys. Rev. 
C77, 061304(R) (2008). 
\bibitem{Klimkiewicz} A.~Klimkiewicz {\em et al.}, Phys. Rev. 
C76, 051603(R) (2007).
\bibitem{Yako} H.~Sagawa {\em et al.}, Phys. Rev. C76, 024301 (2007).
\bibitem{Chen} L.~W.~Chen {\em et al.}, Phys. Rev. Lett. 94, 032701 (2005).
\bibitem{Tsang} M.~B.~Tsang {\em et al.}, Phys. Rev. Lett. 102, 122701 
(2009).
\bibitem{Shetty} D.~V.~Shetty {\em et al.}, Phys. Rev. C76, 024606 (2007).
\bibitem{Warda} M.~Centelles {\em et al.}, Phys. Rev. Lett. 102, 122502 
(2009); M.~Warda {\em et al.}, Phys. Rev. C80, 024316 (2009). 
\bibitem{footnote1} In \cite{Trippa} the correlation is only
between the GDR energy and the symmetry energy at 0.1 fm$^{-3}$: 
however, most of the Skyrme forces which have been proposed 
as acceptable have $L$ below 60 MeV. 
\bibitem{Bender} M.~Bender {\em et al.}, Rev. Mod. Phys. 75, 121 (2003).
\bibitem{Wieland} O.~Wieland {\em et al.}, Phys. Rev. Lett. 
102, 092502 (2009). 
\bibitem{comex3} G.~Col\`o {\em et al.}, Nucl. Phys. A788 (2007) 173c. 
\bibitem{K} S.~Shlomo {\em et al.}, Eur. Phys. J. A30, 
23 (2006); G.~Col\`o, Physics of Particles and Nuclei 39, 286 (2008).
\bibitem{Stone} J.~R.~Stone {\em et al.}, Phys. Rev. C68, 034324 (2003).
\bibitem{Ring01} P.~Ring {\em et al.}, Nucl. Phys. A694, 249 (2001).
\bibitem{Ma02} Z.~Y.~Ma {\em et al.}, Nucl. Phys. A703, 222 (2002).
\bibitem{Cao} Cao Li-gang, Ma Zhong-yu, Chin. Phys. Lett. 25, 1625 (2008).
\bibitem{Sulaksono} A.~Sulaksono {\em et al.}, Phys. Rev. C79, 044306 (2009).  
\bibitem{Alhassid} Y.~Alhassid {\em et al.}, Phys. Rev. Lett. 99, 162504 
(2007); C.~N.~Gilbreth and Y.~Alhassid, private communication.
\bibitem{BSk14} {\tt www-astro.ulb.ac.be/Html/nld\_comb\_ph.html}.
\bibitem{MSk7} {\tt www-nds.ipen.br/RIPL-2/densities.html}.
\bibitem{Danielewicz2} P.~Danielewicz and J.~Lee, Nucl. Phys. A818, 
36 (2009).
\bibitem{Danielewicz1} P.~Danielewicz, Nucl. Phys. A727, 233 (2003).
\bibitem{Vidana} I.~Vida\~na {\em et al.}, Phys. Rev. C80, 045806 (2009).  
\bibitem{Krasznahorkay} A.~Krasznahorkay {\em et al.}, Phys. Rev. Lett. 82, 
3216 (1999). 
\bibitem{Harakeh} M.~N.~Harakeh and A.~van~der~Woude, {\em Giant Resonances} 
(Clarendon Press, Oxford, 2001). 
\end{thebibliography}
\end{document}